\documentclass[
    ,final          
  ]
  {aipproc}

\layoutstyle{6x9}

\begin{document}
\newcommand\doingARLO[2][]{%
  \ifx\mmref\undefined #1\else #2\fi
}
\def\Journal#1#2#3#4{{#1} {\bf #2}, #3 (#4) } 
\def\NCA{\em Nuovo Cimento}
\def\NIM{\em Nucl. Instrum. Methods}
\def\NIMA{{\em Nucl. Instrum. Methods} A}
\def\NPA{{\em Nucl. Phys.} A}
\def\NPB{{\em Nucl. Phys.} B}
\def\PLB{{\em Phys. Lett.}  B}
\def\PRL{\em Phys. Rev. Lett.}
\def\PRD{{\em Phys. Rev.} D}
\def\ZPC{{\em Z. Phys.} C}

\newcommand{\alpi}{\it O(\alpha/\pi)}
\def\beq{\begin{equation}}
\def\eeq{\end{equation}}
\def\lsim{\ ^<\llap{$_\sim$}\ }
\def\gsim{\ ^>\llap{$_\sim$}\ }
\def\r2{\sqrt 2}
\def\beq{\begin{equation}}
\def\eeq{\end{equation}}
\def\beqn{\begin{eqnarray}}
\def\eeqn{\end{eqnarray}}
\def\rmuu{\gamma^{\mu}}
\def\rmud{\gamma_{\mu}}
\def\PL{{1-\gamma_5\over 2}}
\def\PR{{1+\gamma_5\over 2}}
\def\sinW2{\sin^2\theta_W}
\def\AEM{\alpha_{EM}}
\def\mul{M_{\tilde{u} L}^2}
\def\mur{M_{\tilde{u} R}^2}
\def\mdl{M_{\tilde{d} L}^2}
\def\mdr{M_{\tilde{d} R}^2}
\def\mz2{M_{z}^2}
\def\c2b{\cos 2\beta}
\def\au{A_u}
\def\ad{A_d}
\def\cob{\cot \beta}
\def\v#1{v_#1}
\def\tb{\tan\beta}
\def\epem{$e^+e^-$}
\def\KK{$K^0$-$\bar{K^0}$}
\def\wi{\omega_i}
\def\xj{\chi_j}
\def\Wmu{W_\mu}
\def\Wnu{W_\nu}
\def\m#1{{\tilde m}_#1}
\def\mH{m_H}
\def\mw#1{{\tilde m}_{\omega #1}}
\def\mx#1{{\tilde m}_{\chi^{0}_#1}}
\def\mc#1{{\tilde m}_{\chi^{+}_#1}}
\def\mwi{{\tilde m}_{\omega i}}
\def\mxi{{\tilde m}_{\chi^{0}_i}}
\def\mci{{\tilde m}_{\chi^{+}_i}}
\def\mz{M_z}
\def\sw{\sin\theta_W}
\def\cw{\cos\theta_W}
\def\cb{\cos\beta}
\def\sb{\sin\beta}
\def\rwi{r_{\omega i}}
\def\rxj{r_{\chi j}}
\def\rfp{r_f'}
\def\Kik{K_{ik}}
\def\Fq2{F_{2}(q^2)}
\def\tw{\tan\theta_W}
\def\sec2w{sec^2\theta_W}
\def\amu{a_\mu}
\def\maxmodamu{max~|a_\mu^{SUSY}|}
\def\tanbeta{{\rm tan}\beta}
\def\gmin2{(g-2)_\mu}
\def\ecoup{{e \over {{\sqrt2}\sin{\theta_W}}}}
\def\winomass{m_{\tilde W_a}}
\def\zinomass{m_{\tilde Z_{(k)}}}
\def\denom{{\left[4{\nu}^2_{1,2}+{({\mu} \pm {\tilde m}_2)}^2 \right]}
^{-\frac{1}{2}}}
\def\ssq{\sin^2 \delta}
\def\csq{\cos^2 \delta}
\def\BR{B_k^R}
\def\BL{B_k^L}
\def\Gff{G_1(x_{1k})}
\def\Gfs{G_1(x_{2k})}
\def\Gsf{G_2(x_{1k})}
\def\Gss{G_2(x_{2k})}
\catcode`\@=11 
\def\coeff#1#2{{\textstyle{#1\over #2}}}
\def\bra#1{\left\langle #1\right|}
\def\ket#1{\left| #1\right\rangle}
\def\VEV#1{\left\langle #1\right\rangle}
\def\vev#1{\left\langle #1\right\rangle}
\def\lsim{\mathrel{\mathpalette\@versim<}}
\def\gsim{\mathrel{\mathpalette\@versim>}}
\def\@versim#1#2{\vcenter{\offinterlineskip
    \ialign{$\m@th#1\hfil##\hfil$\crcr#2\crcr\sim\crcr } }}
\def\etal{{\em et. al.}}
\def\PL{Phys. Lett.}
\def\PRL{Phys. Rev. Lett.}
\def\PHYREP{Phys. Rep.}
\def\NP{Nucl. Phys.}
\def\PR{Phys. Rev.}
\def\ZPHY{{Z. Phys C} }
\def\NUOVO{Nuovo Cimento}

\title{Theoretical Status of Muon (g-2)}

\author{Utpal Chattopadhyay}{ 
address={Department of Theoretical Physics, Tata Institute of 
Fundamental Research, Homi Bhaba Road, Mumbai 400005, India}
}
\author{Achille Corsetti}{
  address={Department of Physics, Northeastern University, Boston,
  Massachusetts 02115, USA}
}
\author{Pran Nath}{
  address={Department of Physics, Northeastern University, Boston,
  Massachusetts 02115, USA}
}

\begin{abstract}
The theoretical status of the muon anomaly is reviewed including
the recent change in the light by light hadronic correction.
 Specific attention is given to the implications of the shift
 in the difference between the BNL experimental result and the
standard model prediction for sparticle mass limits.
 The implication of the BNL data for Yukawa unification
is discussed and the role of gaugino mass nonuniversalities in the
satisfaction of Yukawa unification is explored.
An analysis of the  BNL constraint for
the satisfaction of the relic density constraint  and for
the search for dark matter is also given.
\end{abstract}

\maketitle


\section{Introduction}  
In this talk we discuss the current status of theory vs 
 experiment for $a_{\mu}=(g_{\mu}-2)/2$ and the implications for new physics.
 Recently a reevaluation of the light by light hadronic contribution to
 $a_{\mu}$ has resulted in a change in the sign of this contribution
 reducing the difference between the BNL experimental result and the standard
 model prediction from $2.6\sigma$ to $1.6\sigma$. 
  In view of this change we reconsider
 the implications for supersymmetry. We carry out the analysis using 
 a $1\sigma$ and a $1.5\sigma$ error corridor around the central value
 of the difference between experiment and theory. For the $1\sigma$
 analysis we find that the upper limits  on sparticle masses remain 
 unchanged from those predicted with the $2.6\sigma$
 difference between experiment and the standard model result with a $2\sigma$
 error corridor. For the $1.5\sigma$ analysis we find that the upper
 limits are substantially increased from the old analysis and the 
 upper limits of the sparticle masses may lie on the
 borderline or beyond of what is accessible at the Large Hadron Collider.
 An important result that arises from the Brookhaven experiment is
 that the sign
 of the $\mu$ parameter is  determined to be positive for a broad class
 of supersymmetric models. However, it is known that Yukawa coupling
 unification typically prefers a negative $\mu$. We discuss a possible way
  out of this problem using nonuniversality of gaugino masses. 
Finally we consider the implications of the Brookhaven result
for neutralino relic density and for the direct detection of 
supersymmetric dark matter in dark  matter detectors.
\section{$g_{\mu}-2$: Experiment vs Standard Model }
Over the last three months the theoretical prediction of $a_{\mu}$ in
the standard model has undergone a significant revision because of 
the change in sign of the light by light (LbL) hadronic 
correction to  $a_{\mu}$.
Thus the previous average for $a_{\mu}^{had}(LbL)$ 
was\cite{hayakawa,bijnens}  
$a_{\mu}^{had}(LbL)=-8.5(2.5)\times 10^{-10}$. However, 
recent reevaluations\cite{knecht,hkrevised,Blokland,Bijnens2,Ramsey}
give a $a_{\mu}^{had}(LbL)$ opposite in sign to the previous
evaluations.  The reevaluations are summarized in Table 1. 
\begin{center} 
\begin{tabular}{|c|c|c|c|}
\multicolumn{2}{c}{Table 1: light by light hadronic correction  } \\
\hline
  authors & $a_{\mu}^{had}(LbL)$ \\
\hline
 Knecht \etal\cite{knecht} & $ 8.3(1.2)\times 10^{-10}$ \\
 \hline
 Hayakawa \& Kinoshita\cite{hkrevised}  & $~8.9(1.5)\times 10^{-10}$ \\
 \hline
 Bijnens \etal\cite{Bijnens2} & $(8.3\pm 3.2)\times 10^{-10}$        \\
 \hline
 Blockland \etal\cite{Blokland} ($\pi^0$) & $5.6\times 10^{-10}$  \\
 \hline
\end{tabular}
\noindent
\end{center}
Now with the old value of the LbL hadronic correction 
the total standard model prediction of 
$a_{\mu}^{SM}=a_{\mu}^{QED}+a_{\mu}^{EW}+a_{\mu}^{had}$
was  $a_{\mu}^{SM}= 11659159.7(6.7)\times 10^{-10}$.
 Using the BNL experimental result\cite{brown} 
of $a_{\mu}^{exp}=11659203(15)\times 10^{-10}$  one finds  
$a_{\mu}^{exp}-a_{\mu}^{SM}= 43(16)\times 10^{-10}$ which gives the
old $2.6\sigma$ deviation between experiment and the standard model.
 However, taking an average of the top three entries in Table 1 for
  $a_{\mu}^{had}(LbL)$ the revised difference between experiment and
  the standard model is 
\begin{equation}
a_{\mu}^{exp}-a_{\mu}^{SM}= 26(16)\times 10^{-10} 
\end{equation}
which is now only a $1.6\sigma $ deviation between experiment and 
 the standard model prediction. [More  recently another evaluation of 
 $a_{\mu}^{had}(LbL)$ based on chiral perturbation theory has been
 given in Ref.\cite{Ramsey} which gives $a_{\mu}^{had}(LbL)
 =(5.5^{+5}_{-6}+3.1\hat C)\times 10^{-10}$ where $\hat C$ stands
 for  unknown low energy constants arising from subleading contributions.
 The authors of  Ref.\cite{Ramsey} view a $\hat C$ range of  $-3$ to 3
 or even larger
 as not unreasonable. The result of Eq.(1) corresponds essentially to a 
 $\hat C=1$ and a much larger value will significantly affect Eq.(1)
 and the conclusions resulting from it.]  Aside from the issue of LbL 
 hadronic correction, the 
remaining part of the hadronic correction contains $\alpha^2$ and
$\alpha^3$ vacuum polarization corrections. In the deduction of Eq.(1) 
we used the evaluation of Ref.\cite{davier} for the $\alpha^2$ correction.  
However, there is 
considerable amount of controversy regarding these 
corrections and this issue is still under 
scrutiny\cite{hadronic}. 
\section{Supersymmetric correction to $g_{\mu}-2$} 
If indeed there is discrepancy between experiment and the standard model
prediction of $a_{\mu}$ then 
it would have important implications
for new physics. Such new physics could be supersymmetry,
compact extra dimensions, muon compositeness,techni-color,
anomalous W couplings,  new gauge bosons, lepto-quarks, or 
radiative muon masses\cite{czar1}. We focus here on supersymmetric models
and specifically 
on supergravity models\cite{msugra} which arise from 
gravity mediated breaking of supersymmetry.
 The soft SUSY breaking sector of the minimal supergravity model 
(mSUGRA) is defined by four parameters: these consist of the universal
scalar mass $m_0$, the universal gaugino mass $m_{1/2}$,
the universal trilinear coupling $A_0$ and $\tan\beta=<H_2>/<H_1>$ 
where $H_2$ gives mass to the up quark and $H_1$ gives mass to the down
quarks and the leptons. 
We will use SUGRA 
models as a benchmark and similar analyses can be carried out 
 in other models such as those based on gauge mediation
and anomaly mediation breaking mechanisms of supersymmetry.
 We begin by discussing the basic one loop contribution to $g_\mu-2$ in
supersymmetry\cite{yuan}. Here the basic contributions are from the chargino 
$\tilde W$ and neutralino $\chi_i$ (i=1,.,4) exchange. 
For the CP conserving case the chargino contribution is the
largest and here one has
\begin{equation}
 a_{\mu}^{\tilde W}={{m^2_\mu} \over {48{\pi}^2}} {{ {A_R^{(a)}}^2} \over
{\winomass^2}}F_1(\left({{m_{\tilde \nu}} \over {m_{\tilde W_a}}}\right)^2)+
{{m_\mu} \over{8{\pi}^2}} {{A_R^{(a)}
A_L^{(a)}} \over {\winomass}} 
F_2(\left({{m_{\tilde \nu}} \over {m_{\tilde W_a}}}\right)^2)
\end{equation}
where
$A_L(A_R)$ are the left(right) chiral amplitudes and are defined by
\begin{eqnarray}
A_R^{(1)}=-{\ecoup}\cos {\gamma_1}; \quad A_L^{(1)}={(-1)^\theta}
{{{em_\mu}\cos{\gamma_2}} \over {{2}M_W\sin\theta_W \cos {\beta}}}\nonumber\\
A_R^{(2)}=-{\ecoup}\sin {\gamma_1}; \quad A_L^{(2)}=-
{{{e m_\mu}\sin{\gamma_2}} \over
{{2}M_W \sin\theta_W\cos {\beta}}}
\end{eqnarray}
and where $\gamma_i$ are the mixing angles and $F_1,F_2$ are form
factors.
Recently, the absolute signs of the supersymmetric contribution was 
checked by taking the supersymmetric limit\cite{in}. 
There are some interesting features of the SUSY contribution.  
One finds that since $ A_L\sim 1/\cos\beta$ one has\cite{lopez,chatto}
$\amu^{SUSY}\sim \tan\beta$. Further, it is easy to show that the sign
of $\amu^{SUSY}$ is correlated with the sign of $\mu$\cite{lopez,chatto}.
 Thus one finds that $\amu^{SUSY}>0$ for  $\mu>0$ and  $\amu^{SUSY}<0$
 for $\mu<0$ where we use the sign convention of Ref.\cite{sugra}.
\section{ Implications of Data}
\paragraph{Upper limits on sparticle masses}  
Soon after the BNL result became available\cite{brown} 
a large number of analyses
appeared in the literatures exploring the implications of the results
for new physics\cite{everett}. These analyses were based on the result
 $a_{\mu}^{exp}-a_{\mu}^{SM}=43(16)\times 10^{-10}$ which as is now
 realized is based on using the wrong sign of the light by light hadronic
 correction. Using the above result and a the 2$\sigma$ error corridor 
so that $ 10.6\times 10^{-10}<$$ a_{\mu}^{SUSY}$$<76.2\times 10^{-10}$
the BNL data leads to the following sparticle mass limits 
in mSUGRA\cite{chatto2}: 
$m_{\tilde W}\leq 650 {~\rm GeV},~m_{\tilde \nu} \leq 1.5 {~\rm TeV}~
(\tan\beta\leq 55)$ and 
$m_{1/2}\leq 800 {~\rm GeV},~m_0\leq 1.5 {~\rm TeV}~(\tan\beta\leq 55) $.
Since the LHC can explore squarks/gluinos up 
to 2 TeV  the BNL result implies
that sparticles should become visible at the LHC\cite{cms}. 
\begin{figure}
  \includegraphics[height=.3\textheight]{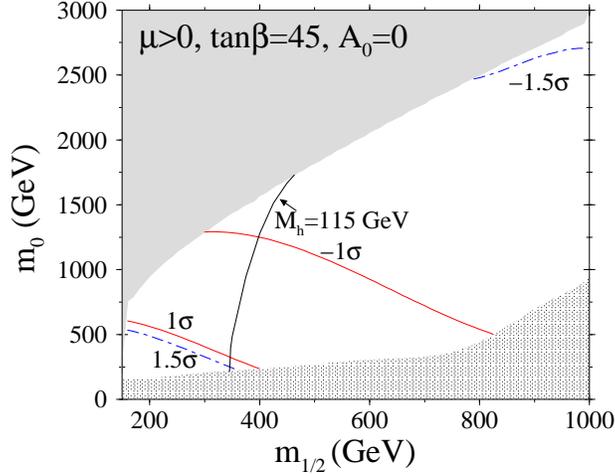}
  \caption{Upper and lower limits in $m_0-m_{1/2}$ plane 
   corresponding to $1\sigma$ and $1.5\sigma$  
  constraints from $a_{\mu}^{SUSY}$  for $\tan\beta=45$.
   The top left gray region does not 
satisfy the radiative electroweak symmetry breaking requirement while
 the bottom patterned region near the 
higher $m_{1/2}$ side and on the border of the white allowed regions is
discarded either because of stau or the CP-odd Higgs boson turning
 tachyonic at the tree level (from Ref.\cite{ccnyuk}).}
\end{figure}
Next we assess the situation as a consequence of the change in the sign of
the light by light hadronic correction which results in Eq.(1). 
In this case a $2\sigma$ error corridor would not lead
to upper limits for the sparticle masses. However, one can get
interesting constraints if one imposes a  $1\sigma$ or a $1.5\sigma$
constraint. A $1\sigma$ constraint actually yields exactly the same 
upper limits as before so in this case the analysis of 
Ref.\cite{chatto2} remains 
valid as far as the upper limits are concerned. The case of $1.5\sigma$
was analyzed in Ref.\cite{ccnyuk}
 and it was found, as expected, that the upper limits
go up considerably. In Fig.(1) results are presented for the case of 
$\tan\beta=45$. Here one finds that the upper limits on 
$m_0$ and $m_{{1}/{2}}$ are 
considerably larger than for the analysis of Ref.\cite{chatto2}. 
Specifically one finds from Fig.(1) that the upper limit on $m_0$ 
in the range of the parameter space exhibited already exceeds 2.5 TeV 
which is on the borderline of the reach of the LHC.
The upper limits for other values of $\tan\beta$ are sharply dependent
on the value of $\tan\beta$. A more complete analysis of the constraint
for the $1.5\sigma$ case can be found in Ref.\cite{ccnyuk}. 

\noindent
Another interesting implication of the BNL result is that
under the assumption of CP conservation and setting $a_{\mu}^{SUSY}=
$$a_{\mu}^{exp}-a_{\mu}^{SM}$ the BNL data determines 
$sign(\mu)=+1$ (see, e.g., Ref.\cite{chatto2}).
It known that $\mu>0$ is favored by the 
$b\rightarrow s+\gamma$ constraint\cite{bsgamma,bsgammanew} and also 
 favored for dark matter searches. The implications of $\mu>0$ for
 dark matter will be discussed in
 further detail below. One issue of concern relates to the possibility
 that the  supersymmetric effects may be masked by effects arising from
  low lying extra dimensions.
This possibility was examined in Ref.\cite{ny1}
  in a model with one large extra
dimension compactified on $S^1/Z_2$ with radius R ($M_R=1/R=O({~\rm TeV})$).
The extra dimension contributes to the Fermi constant and by a 
comparison of the standard model prediction with the experimental value
of the Fermi constant\cite{ny1} one can place a limit on the extra dimension of about
$M_R>3$ TeV. With this size value of $M_R$ one finds that the contribution 
of the extra dimension to $a_{\mu}$ is negligible\cite{ny1}
compared to the supersymmetric
contribution.  For the case of strong gravity the effect on $a_{\mu}$ from the
Kaluza-Klein excitations of the graviton in the case d=2 is 
also small\cite{graesser}
since here the fundamental Planck scale  $M_*$ is found to have a lower
limit of $M_* >3.5$ TeV from the recent gravity experiment\cite{gravity}. 
The above
exhibits the fact that $g_{\mu}-2$ is not an efficient probe of extra 
dimensions. Other techniques such as energetic dileptonic signals at 
LHC would
be more efficient signals for the exploration of extra dimensions\cite{nyy}.
 An important effect that can modify the supersymmetric 
contribution is the effect of CP violating phases. This topic has been analyzed in
several works\cite{in}. Specifically it is found that the  BNL data can be used
to constrain the CP phases and strong constraint on the phases are 
found to exist\cite{icn}.

\paragraph{Positivity of $\mu$ and Yukawa Unification} 
 We discuss now another aspect of the $g_\mu-2$ constraint and this 
concerns Yukawa unification in supersymmetric models. It has been known
for some time that  $b-\tau$ Yukawa coupling unification 
typically prefers a negative $\mu$\cite{bagger,deboer}. 
Thus the positively of the $\mu$ sign
implied by the BNL data appears a priori to pose a problem for 
Yukawa unification.  Now the reason why Yukawa unification typically
prefers 
a negative $\mu$ is easily understood from the fact that  
such unification requires a negative supersymmetric correction to the 
b quark mass and a negative correction to the b quark mass is easily 
obtained when $\mu$ is negative. To illustrate this phenomenon we
consider the gluino and chargino exchanges which contribute the  
 largest supersymmetric correction to the b quark mass.
 Thus one has\cite{susybtmass}
$\Delta_b^{\tilde g}= \frac{2\alpha_3\mu M_{\tilde g}}{3\pi}   
\tan\beta I(m_{\tilde b_1}^2, m_{\tilde b_2}^2,M_{\tilde g}^2)$
 and
$\Delta_b^{\tilde \chi^+}= \frac{Y_t\mu A_t}
{4\pi} \tan\beta I(m_{\tilde t_1}^2, m_{\tilde t_2}^2,\mu^2)$
where $Y_t=\lambda_t^2/4\pi$ where $I>0$.
Generally the gluino exchange contribution tends to be the larger one
and here one finds that for a positive $M_{\tilde g}$ which is typically
the case  a negative $\mu$ indeed leads to a negative 
correction to the b quark mass which in turn leads to the usual result that 
$b-\tau$ unification prefers a negative $\mu$.
\begin{figure}
  \includegraphics[height=.3\textheight]{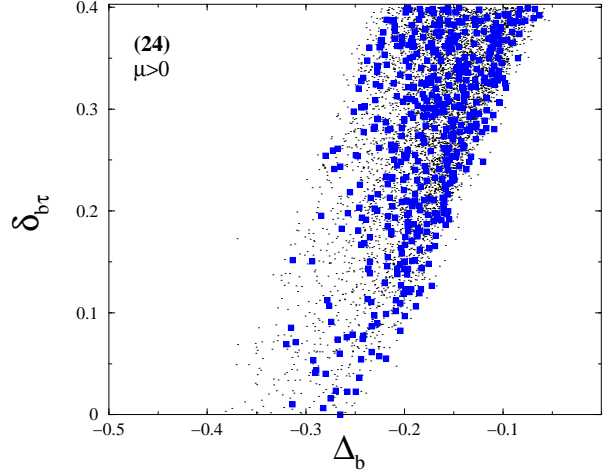}
  \caption{$\delta_{b\tau}$ vs $\Delta_b$, the SUSY correction to $m_b$,
   for the 24-plet case, when $\tan\beta<55$,
$0<m_0<2$ TeV, $-1 ~{\rm TeV} <C_{24}m_{1/2}<1 ~{\rm TeV}$,
$-6 ~{\rm TeV}<A_0<6~{\rm TeV}$ and $\mu>0$, where $C_{24}$ is as
defined in Ref.\cite{cnbtau}.
The dots refer to $b-\tau$ unification at the shown level 
and filled squares additionally
represent points which satisfy both the $b \rightarrow s+\gamma$ and 
the muon $g-2$ constraints (from Ref.\cite{cnbtau}).}
\end{figure} 
There are several solutions discussed 
recently to overcome this problem\cite{bf,bdr,ky,cnbtau}. 
One simple possibility discussed
in Refs.\cite{ky,cnbtau} is that of nonuniversal gaugino masses where 
the sign of the
gluino mass is negative relative to the mass of the SU(2) gaugino mass.
In this case one can obtain a negative contribution to the 
b quark mass while maintaining a positive $\mu$. 
 The basic idea is that with nonuniversalities one can have
the sign of $SU(2)$ and $SU(3)$ gauginos to be opposite. 
A positive positive $\mu$ and a positive  $\tilde m_2$ are 
 consistent with the
BNL data, while a  positive $\mu$ and a  negative $\tilde m_3$
 gives a negative correction
to the b qaurk mass and leads to $b-\tau$ unification.
We considered two classes of models, one based on SU(5) and the 
other on SO(10). For the SU(5) case one has that the the gaugino
 mass matrix
can transform like the symmetric product of $(24\times 24)_{sym}$ which 
has the expansion of $1+24+75+200$. In this case the gaugino masses arising
from the 24 plet has the opposite sign between the SU(2) 
and SU(3) gauginos\cite{anderson}.
Thus one  can choose $\mu$ positive and the gluino mass to be negative
which gives a negative contribution to the b quark mass and allows for
the satisfaction of the $b-\tau$ unification constraints.
The degree of unification defined by 
$\delta_{b\tau}=(|\lambda_b-\lambda_{\tau}|)/\lambda_{\tau}$ vs the 
correction to the b quark mass is exhibited in Fig.(2) for the 24 plet
case with a positive $\mu$ sign. One finds that ${b-\tau}$ unification
 can be satisfied to a high degree of accuracy with an appropriate negative 
correction $\Delta_b$ to the b quark mass. A similar analysis holds
for the SO(10) case. Here the gaugino mass matrix can transform like
symmetric product of $(45\times 45)_{sym}$ which 
has the expansion of $1+54+210+770$. Here for the case when the symmetry
breaking occurs pattern is of the form 
$SO(10)\rightarrow SU(4)$$\times SU(2)\times SU(2)$
$\rightarrow SU(3)\times SU(2)\times U(1)$ one finds that the 
$SU(3), SU(2), U(1)$ gaugino masses arising from the 54 plet 
are in the ratio\cite{chamoun} $M_3:M_2:M_1$$=1:-3/2:-1$. However, for the 
symmetry pattern 
$SO(10)\rightarrow SU(2)$$\times SO(7)$
$\rightarrow SU(3)\times SU(2)\times U(1)$ one finds that the 
$SU(3), SU(2)$, and $U(1)$ gaugino masses arising from the 54 plet 
are in the ratio\cite{chamoun} $M_3:M_2:M_1$$=1:-7/3:1$. 
We will call this case $54'$.
An analysis similar to that for the 24 plet case can be carried out
for the 54 and $54'$ cases and
one finds that $b-\tau$ unification occurs once again for these cases.  
 
\paragraph{Implications for Relic Density and Dark Matter Search}  
As  noted earlier the BNL data indicates a positive value of 
$\mu$ for mSUGRA. This result has important implications for dark 
matter. As already indicated 
 a positive $\mu$ is preferred by the constraint
imposed by the flavor changing neutral current process 
$b\rightarrow s+\gamma$\cite{bsgamma} in that the experimental value for
 this branching
ratio imposes severe constraints on the SUSY parameter space for the
negative sign of $\mu$ but imposes much less stringent constraints 
on the parameter space for a positive value of $\mu$. Thus a positive $\mu$
is more favorable for supersymmetric dark matter analysis in 
that it allows for a large amount of the parameter space 
of the model where relic density constraints can be satisfied along with 
satisfying the $b\rightarrow s+\gamma$ constraint.  
Furthermore, it also turns out that 
detection rates for a positive $\mu$ are generally larger than for
a negative $\mu$. 
Thus after the BNL data became available it was immediately realized 
that the positive $\mu$ sign indicated by the BNL data was favorable
for dark matter searches\cite{chatto2,everett}. More detailed analyses
 were done in several 
subsequent works and the parameter space of mSUGRA was further constrained
from the relic density constraints. Now another way that the BNL data 
constrains dark matter is through the Yukawa unification conditions.
Here we discuss the implications of this constraint on dark matter.
As discussed in the section on $b-\tau$ unification above, one  finds that
this unification can be achieved  with a positive $\mu$ in a variety of ways.
One possibility discussed above arose from nonuniversal gaugino masses.
We discussed two main scenarios for nonuniversalities corresponding to the
SU(5) and SO(10) cases. For SU(5) the gaugino mass nonuniversalities  
arising from the 24 plet case allows a negative contribution to the 
b quark mass with a positive $\mu$ and leads to regions of the parameter
space where $b-\tau$ unification occurs. Analysis in this region
exhibits that all of the spectrum lies within the usual naturalness 
limits\cite{ccn}.
 It is interesting to investigate
if this region of the parameter space also leads to a satisfaction of the
relic density constraint. We consider a rather liberal corridor here
corresponding to the range $0.02\leq \Omega h^2 \leq 0.3$. The analysis
shows that significant regions of the parameter space exist where these
constraints are satisfied. An analysis of the detection rates in the 
direct detection of dark matter was also given\cite{ccnyuk}.
One finds\cite{ccnyuk} that the detection rates lie in a range that 
can be fully explored in the new generation of experiments currently
underway and those which are planned in the 
future (see, e.g., Ref.\cite{genius}). A similar analysis can 
be carried out for the SO(10) case. The sparticle masses consistent with
the BNL $1\sigma$ constraint as given by Eq.(1), consistant with 
the $b\rightarrow s+\gamma$
constraint and consistent with the relic constraint are given in Table 2
 for the
24 plet case of SU(5) and for the 54 and $54'$ cases of SO(10).
 \begin{center}
\begin{tabular}{|c|c|c|c|}
\multicolumn{4}{c}{Table 2: ~Sparticle mass ranges for 24, 54, and $54'$
 cases from Ref.\cite{ccnyuk}} \\
\hline
  Particle & {\bf 24} (GeV) & {\bf 54} (GeV) & {\bf $54'$} (GeV)\\
\hline
  $\chi_1^0$ & 32.3 - 75.2 & 32.3 - 81.0 & 32.3 - 33.4\\
\hline
  $\chi_1^\pm$ & 86.9 - 422.6 & 94.6 - 240.8 & 145.8 - 153.9\\
  \hline
  $\tilde g$ & 479.5 - 1077.2 & 232.5 - 580.3 & 229.8 - 237.4 \\
 \hline
  $\tilde\mu_1$ & 299.7 - 1295.9 & 480.5 - 1536.8 & 813.1 - 1196.3\\
 \hline
  $\tilde\tau_1$ & 203.5 - 1045.1 & 294.2 - 1172.6 & 579.4 - 863.7\\
\hline
  $\tilde u_1$ & 533.6 - 1407.2 & 566.7 - 1506.4 & 822.9 - 1199.8\\
 \hline
  $\tilde d_1$ & 535.1 - 1407.5 & 580.3 - 1546.2 & 845.1 - 1232.5\\
 \hline
  $\tilde t_1$ & 369.9 - 975.2 & 271.5 - 999.6 & 513.7 - 819.9\\
 \hline
  $\tilde b_1$ & 488.2 - 1152.8 & 158.1 - 1042.0 & 453.2 - 749.9\\
  \hline
  $h$ & 104.3 - 114.3 & 103.8 - 113.3 & 108.1 - 110.9\\
 \hline
 \end{tabular}\\
\end{center}
 Here one
finds some interesting features in the spectrum. Thus in these scenarios
the neturalino mass lies below 81 GeV and the higgs boson mass lies 
below 115 GeV in the three scenarios considered in Table 2.
 The Higgs mass ranges of Table 2 are
consistent with the current Higgs mass limits from LEP\cite{lephiggs}
taking into account the $\tan\beta$ dependence\cite{Sopczak}.
Further these mass ranges can be fully explored
in RUNII of the Tevatron. Similarly the mass ranges of the other 
sparticle masses of Table 2 can be explored in RUNII of the Tevatron
via the trileptonic signal\cite{trilep} and other 
techniques\cite{sugra} while the full range for most of the spectrum 
of Table 2 can be explored at the LHC\cite{cms}. 
\section{Conclusion}
 There is a significant amount of more data from the 2000 runs and BNL 
 eventually hopes to measure $a_{\mu}$ to an accuracy of $4\times 10^{-10}$.
 On the other hand, reanalyses of the hadronic correction are still 
 underway to pin down further the size of these corrections. 
If the deviation between the central value of 
experiment and the standard model prediction persists at the current level 
but the error  is significantly reduced one could still see 
a possibility of approaching the discovery limit. 
Needless to say the implications of a sizable deviation between 
experiment and theory are enormous as forseen in early works\cite{yuan}
and elucidated further in several subsequent 
works\cite{lopez,chatto,everett,chatto2}. 
Specifically, the light Higgs boson should show up in RUNII of the 
Tevatron and 
most of the sparticles ($\tilde g, \tilde q, \tilde W,$ etc)
should become visible at the LHC. Further, a positive $\mu$ sign
implied by the BNL data along with a low lying sparticle spectrum 
is very encouraging for the search for supersymmetric dark matter. 

\begin{theacknowledgments}
This work was  supported in part by NSF grant PHY-9901057.
\end{theacknowledgments}

\end{document}